\documentclass[fleqn,usenatbib,useAMS]{mnras}

\usepackage{graphicx}	
\usepackage{amsmath}	
\usepackage{amssymb}	
\usepackage{multicol}        
\usepackage{bm}		
\usepackage{pdflscape}	
\raggedcolumns

\usepackage[T1]{fontenc}
\usepackage{ae,aecompl}
\usepackage{ulem}

\usepackage{newtxtext,newtxmath}

\title[Distance to MAXI J1820+070]{A radio parallax to the black hole X-ray binary MAXI\,J1820+070}

\author[P. Atri]{P. Atri$^{1}$ ,\thanks{Email: pikky.atri@icrar.org (PA)}
J. C. A.~Miller-Jones$^{1}$ ,
A. Bahramian $^{1}$ ,
R. M. Plotkin $^{1,2}$ ,
A. T. Deller$^{3}$ ,
\newauthor 
P. G. Jonker$^{4,5}$ ,
T. J. Maccarone$^{6}$,
G. R. Sivakoff$^{7}$,
R. Soria$^{8,9}$ ,
D. Altamirano$^{10}$ ,
\newauthor
T. Belloni$^{11}$ ,
R. Fender$^{12}$ ,
E. Koerding$^{13}$ ,
D. Maitra$^{14}$ ,
S. Markoff$^{15}$ ,
S. Migliari$^{16,17}$ ,
\newauthor
D. Russell$^{18}$ ,
T. Russell$^{15}$ ,
C. L. Sarazin$^{19}$ ,
A. J. Tetarenko$^{20}$ ,
V. Tudose $^{21}$
\\
$^{1}$International Centre for Radio Astronomy Research, Curtin University, GPO Box U1987, Perth, WA 6845, Australia\\
$^{2}$Department of Physics, University of Nevada, Reno, NV 89557, USA\\
$^{3}$Centre for Astrophysics and Supercomputing, Swinburne University of Technology, Mail Number H11, P.O. Box 218, Hawthorn, VIC 3122 Australia\\
$^{4}$SRON, Netherlands Institute for Space Research, Sorbonnelaan 2, NL-3584 CA Utrecht, the Netherlands\\
$^{5}$Department of Astrophysics/IMAPP, Radboud University Nijmegen, POBox 9010, NL-6500 GL, Nijmegen, the Netherlands\\
$^{6}$Department of Physics, Box 41051, Science Building, Texas Tech University, Lubbock, TX 79409-1051, USA\\
$^{7}$Department of Physics, University of Alberta, CCIS 4-183, Edmonton, AB T6G 2E1, Canada\\
$^{8}$School of Astronomy and Space Sciences, University of Chinese Academy of Sciences, Beijing, China\\
$^{9}$Sydney Institute for Astronomy, School of Physics A28, The University of Sydney, Sydney, New South Wales, Australia\\
$^{10}$Physics and Astronomy, University of Southampton, Southampton, Hampshire SO17 1BJ, UK\\
$^{11}$INAF - Osservatorio Astronomico di Brera, Via E. Bianchi 46, I-23807 Merate, Italy\\
$^{12}$Astrophysics, Department of Physics, University of Oxford, Keble Road, Oxford OX1 3RH, UK\\
$^{13}$Department of Astrophysics/IMAPP, Radboud University, Nijmegen, PO Box 9010, 6500 GL
Nijmegen, the Netherlands\\
$^{14}$Department of Physics and Astronomy, Wheaton College, Norton, MA 02766, USA\\
$^{15}$Astronomical Institute 'Anton Pannekoek', University of Amsterdam, Science Park 904, NL-1098XH Amsterdam, the Netherlands\\
$^{16}$XMM-Newton Science Operations Centre, ESAC/ESA, Camino Bajo del Castillo s/n, Urb. Villafranca del Castillo, 28691 Villanueva de la Ca\~nada, Madrid, Spain\\
$^{17}$Institute of Cosmos Sciences, University of Barcelona, Mart\'i i Franqu\`es 1, 08028 Barcelona, Spain\\
$^{18}$Center for Astro, Particle and Planetary Physics, New York University Abu Dhabi, PO Box 129188, Abu Dhabi, UAE\\
$^{19}$Department of Astronomy, University of Virginia, 530 McCormick Road, Charlottesville, VA 22904, USA\\
$^{20}$East Asian Observatory, 660 N. A\={o}h\={o}ku Place, University Park, Hilo, Hawaii 96720, USA\\
$^{21}$Institute for Space Sciences, Atomistilor 409, PO Box MG-23, 077125 Bucharest-Magurele, Romania}
\date{Accepted XXX. Received YYY; in original form ZZZ}

\pubyear{2019}

\begin{document}
\label{firstpage}
\pagerange{\pageref{firstpage}--\pageref{lastpage}}
\maketitle

\begin{abstract}
Using the Very Long Baseline Array and the European Very Long Baseline Interferometry Network we have made a precise measurement of the radio parallax of the black hole X-ray binary MAXI\,J1820+070, providing a model-independent distance to the source. Our parallax measurement of ($0.348\pm0.033$)\,mas for MAXI\,J1820+070 translates to a distance of ($2.96\pm0.33$)\,kpc. This distance implies that the source reached ($15\pm3$)\,per cent of the Eddington luminosity at the peak of its outburst. Further, we use this distance to refine previous estimates of the jet inclination angle, jet velocity and the mass of the black hole in MAXI J1820+070 to be ($63\pm3$)$^{\circ}$, ($0.89\pm0.09$)\,$c$ and ($9.2\pm1.3$) $M_{\odot}$, respectively.
\end{abstract}




\section{Introduction}
The distances to astronomical systems are essential quantities that allow us to draw a connection between observed and physical parameters (e.g., flux and luminosity, angular and physical speeds). In particular, for black hole X-ray binaries (BHXBs) the relation between different accretion phases and Eddington luminosity fractions can then be studied \citep{Maccarone2003}, and the systems can be accurately placed on the X-ray/radio luminosity plane \citep{Gallo2018} to probe jet/accretion coupling. 
With accurate distances, we can break the degeneracy between the inclination angles of radio jets and their speeds \citep{Fender2003}. A well constrained inclination angle can also help in obtaining accurate mass estimates of the black hole (BH) from the mass function of the system \citep{Cantrell2010}. An accurate distance helps in obtaining a well constrained three dimensional space velocity and hence the potential kick velocity (PKV) distribution of the system \citep{Atri2019}, providing clues as to the BH formation mechanism \citep{Nelemans1999}.\par
Distances to BHXBs are often estimated by studying the optical/infrared spectra of the donor star \citep{Jonker2004}.
The proper motion of the receding and approaching jets can also be used to place an upper limit on the distance of the BHXBs \citep{Mirabel1994}. Lower limits on the distance can be estimated by measuring either the interstellar extinction \citep{Zdziarski1998,Zdziarski2004} or the velocities of H\,{\sc i} absorption \citep[e.g.][]{Chauhan2019} towards the source. X-ray dust scattering halo has also been used to constrain the distance to some the BHXBs \citep[e.g.][]{Xiang2011}. However, these methods are all model-dependent, involving certain assumptions. The only model-independent method of distance determination is measuring a high significance trigonometric parallax. However, given their typical distances (several kpc), such high-precision observations can only be carried out using Very Long Baseline Interferometry (VLBI) or by satellites like \textit{Gaia} \citep{Brown2018,Gandhi2019}. \textit{Gaia}'s capabilities to conduct high precision astrometry of BHXBs in the Galactic plane can be limited due to high extinction and low optical brightness outside of outburst. \textit{Gaia} also has a global zeropoint offset in its parallax measurements, the magnitude of which is still under debate \citep{Chan2019}. Therefore, targeted VLBI astrometry campaigns on outbursting BHXBs remain crucial. Of the $\sim$60 BHXB candidates \citep{Tetarenko2016a}, only V404\,Cyg \citep{Miller-Jones2009b}, Cyg\,X--1 \citep{Reid2011,Brown2018} and GRS\,1915+105 \citep{Reid2014} have a directly measured parallax at >5$\sigma$ significance. \par
In March 2018 (MJD 58188), the BHXB MAXI\,J1820+070 (ASASSN-18ey, hereafter MAXI\,J1820) went into outburst and was detected as a bright X-ray \citep{Kawamuro2018} and optical \citep{Tucker2018} transient. Archival photographic plates show two past outbursts in 1898 and 1934 \citep{Grindlay2012}, suggesting an outburst recurrence timescale of $\sim$40 years for this source \citep{Kojiguchi2019}. In its March 2018 outburst, the transient made a complete hard (until MJD 58303.5) to soft (MJD 58310.7--58380.0) to hard state (from MJD 58393.0) outburst cycle \citep{Shidatsu2019}. Radio observations during and after the transition from hard to soft state revealed an apparently superluminal approaching jet (Bright et al. submitted). The source almost faded to quiescence in February 2018 \citep{Russell2019}, enabling optical spectroscopic studies \citep{Torres2019} to dynamically confirm the presence of a BH. It subsequently showed two reflares beginning on MJD 58552 and MJD 58691, which each lasted a couple of months \citep{Ulowetz2019,Zampieri2019}. \par
The PKV distribution of MAXI\,J1820 is loosely constrained as it was based on the \textit{Gaia} proper motion measurement $\mu_{\alpha}\cos\delta = -3.14\pm0.19$\,mas\,yr$^{-1}$ and $\mu_{\delta} = -5.9\pm0.22$\,mas\,yr$^{-1}$ and broad limits on the distance \citep[1.7$-$3.9\,kpc;][]{Atri2019}. The distance inferred from the \textit{Gaia} parallax depended heavily on the priors used to invert the parallax \citep[2.1--7.2\,kpc;][]{Gandhi2019}. Thus, to obtain a more accurate parallax of MAXI\,J1820 we carried out a targeted VLBI campaign while the source was in the hard X-ray spectral state, emitting steady compact radio jets. 

\section{Observations and Data reduction}\label{Observations}
\begin{table*}
        \begin{center}
                \caption{Summary of the observations of MAXI\,J1820+070. The observing bandwidth was 32\,MHz for both the VLBA and the EVN. We give the measured target positions for each epoch. When both J1813 and J1821 were observed as phase calibrators, we give target positions phase referenced to 1821. The error bars denote the statistical errors in the position measurement of the targets. \label{tab:table2}}
                \scalebox{0.8}{%
                \begin{tabular}{l c c c c c l l l}
                \hline \hline
                        Project Code & MJD & Array & Frequency &  Phase Calibrator & Check source & RA (J2000) & Dec (J2000) & Peak Intensity \\
                        &  &  & (GHz) & & & (18$\rm{^h}$20$\rm{^m}$) & (07$^\circ$11$^\prime$ ) & (mJy\,bm$^{-1}$)\\
                        \hline
                        BM467A &  58193.65 & VLBA & 15 & J1813        & J1821        & 21$\fs$9386536(1)  & 07$\farcs$170025(4) & 31.6$\pm$0.2\\
                        BM467O &  58397.01 & VLBA & 15 & J1813        & J1821        & 21$\fs$9384875(4)  & 07$\farcs$166302(10)  & 5.84$\pm$0.08\\
                        EA062A &  58407.71 & EVN  & 5  & J1821        & J1813        & 21$\fs$9384883(33) & 07$\farcs$166075(27)  & 1.55$\pm$0.05\\
                        BM467R &  58441.73 & VLBA & 15 & J1813        & J1821        & 21$\fs$9384770(9) & 07$\farcs$165549(31)  & 0.56$\pm$0.03\\
                        EA062B &  58457.04 & EVN  & 5  & J1821        & J1813        & 21$\fs$938437(16)  & 07$\farcs$16485(12)   & 0.16$\pm$0.03\\
                        BM467S &  58474.86 & VLBA & 5  &  J1821, J1813 & J1813, J1821 & 21$\fs$938462(14)  & 07$\farcs$16498(41)   & 0.13$\pm$0.02\\
                        EA062C &  58562.25 & EVN  & 5  & J1821        & J1813        & 21$\fs$9384324(12) & 07$\farcs$163533(10)  & 3.86$\pm$0.05\\
                        BA130B &  58718.06 & VLBA & 5  & J1821, J1813 & J1813, J1821 & 21$\fs$9382958(8)  & 07$\farcs$160872(21)  & 4.09$\pm$0.05\\
                               &  58718.14 & VLBA & 15 & J1821, J1813 & J1813, J1821 & 21$\fs$9383011(3)  & 07$\farcs$160709(14)  & 4.32$\pm$0.12\\
                        BA130C &  58755.04 & VLBA & 5  & J1821, J1813 & J1813, J1821 & 21$\fs$9382761(28) & 07$\farcs$159845(93)  & 1.00$\pm$0.05\\
                               &  58755.12 & VLBA & 15 & J1821, J1813 & J1813, J1821 & 21$\fs$9382730(22) & 07$\farcs$160090(74)  & 1.20$\pm$0.12\\
                        \hline
                \end{tabular}
                }
        \end{center}
\end{table*}
We monitored MAXI\,J1820 (see Table \ref{tab:table2}) with the Very Long Baseline Array (VLBA), using both the Jet Acceleration and Collimation Probe of Transient X-ray Binaries (JACPOT-XRB) program \citep[e.g.,][proposal code BM467]{Miller-Jones2012,Miller-Jones2019}, and a long-running astrometry program \citep[e.g.,][proposal code BA130]{Atri2019}, supplemented with a targeted European VLBI Network (EVN) campaign (proposal code EA062). All observations used a combination of J1821+0549 (J1821 hereon) and J1813+065 (J1813 hereon) as the phase reference and astrometric check sources. The calibrator positions were taken from the Radio Fundamental catalogue (rfc2015a\footnote{\url{http://astrogeo.org/vlbi/solutions/rfc_2015a/rfc_2015a_cat.html}}). Our assumed positions (J2000) were RA = 18$\rm{^h}$21$\rm{^m}$27\fs305837, Dec = 05$^\circ$49$^\prime$10\farcs65156 for J1821, and  RA = 18$\rm{^h}$13$\rm{^m}$33\fs411619, Dec = 06$^\circ$15$^\prime$42\farcs03366 for MAXI J1813. We imaged the calibrated data and determined the target position by fitting a point source in the image plane for every epoch.

\subsection{VLBA campaigns}\label{vlba_data}
In the hard state at the start and end of an outburst, the radio jets are compact and ideal for astrometry. We took four epochs (March, October, November and December 2018) of observations via the JACPOT-XRB program in which the first three epochs (proposal codes BM467A, BM467O and BM467R) were observed at 15\,GHz. These observations used J1813 (1.93$^\circ$ away from MAXI\,J1820) as the phase calibrator (being brighter at 15\,GHz), and cycled every $\sim 2$\,min between it and the target, observing the check source J1821 (1.39$^\circ$ away from MAXI\,J1820) once every eight cycles. We observed geodetic blocks \citep{Reid2009} for half an hour at the beginning and end of each observation to better model the troposphere. The data were correlated using DiFX \citep{Deller2011} and standard calibration steps were followed using the Astronomical Image Processing System \citep[AIPS 31DEC17;][]{Greisen2003}. As the source faded in December 2018, we took the final epoch in the more sensitive 5\, GHz band with the observing scheme J1821--J1813--MAXI\,J1820.\par
We also observed under the astrometry program BA130 during its August 2019 reflare. We cycled through all these sources at 5\,GHz for 1.5\,hrs, and then at 15\,GHz for 2.5\,hrs, with geodetic blocks at the start and end of the observation. We followed standard calibration techniques within \textsc{AIPS} (31DEC17). MAXI\,J1820 was phase referenced separately to J1821 and J1813. Data from the Mauna kea dish was removed for the complete duration of BA130B due to very high dispersive delays that were calculated by the task TECOR. 
\subsection{Parallax campaign: EVN data}\label{evn_parallax}
We were also approved to observe with the EVN at 5\,GHz during the expected peaks of parallax offset in RA in the months of March and September, and in declination in the months of June and December. The source was observed in October 2018, December 2018 and March 2019, but faded below the detection capability of the EVN before the scheduled June 2019 epoch. J1821 was chosen as the phase reference source due to its close proximity to MAXI\,J1820 and was observed every 4\,minutes, with J1813 used as a check source. The data were reduced using AIPS (31DEC17). We used the bandpass, a-priori amplitude and parallactic angle corrections produced by the EVN pipeline, and corrected for ionospheric dispersive delays. We then performed phase, delay and rate calibration using standard procedures in \textsc{AIPS} and extrapolated the phase and delay solutions derived from J1821 to both J1813 and MAXI\,J1820.
\subsection{Mitigating systematic astrometric biases}\label{systematics}
To mitigate against systematic errors arising from low elevations, we removed all data below 20$^\circ$ in elevation. We also removed half an hour of data around the sunrise/sunset times at each station when the ionosphere could be changing rapidly.  To prevent source structure sampled with differing {\it uv}-coverage from affecting our astrometry, we made global models of the two phase reference sources by stacking all EVN data at 5\,GHz, and all VLBA data at each of 5 and 15\,GHz.  These global models were used to derive our final phase and delay solutions for each epoch, which were then extrapolated to MAXI\,J1820 and the relevant check source.\par
The JACPOT-XRB program (see \S~\ref{vlba_data}) and the EVN parallax campaign (see \S~\ref{evn_parallax}) observed different extragalactic sources as phase calibrators. Hence, we scheduled the BHXB astrometry program observations (BA130B and BA130C, see \S~\ref{vlba_data}) such that MAXI\,J1820 could be independently phase referenced to both J1813 and J1821. Combining these two observations with BM467S (which also cycled between all three sources), we measured an average shift of +0.29$\pm$0.08\,mas in RA and +0.05$\pm$0.02\,mas in Dec in the position of MAXI\,J1820 when it was phase referenced to J1813 as compared to when it was phase referenced to J1821. TO account for this astrometric frame shift, we shifted the target positions measured when phase referenced to J1813 by $-$0.29$\pm$0.08\,mas and $-$0.05$\pm$0.02\,mas in RA and Dec, respectively.\par
To estimate the systematics arising from the troposphere ($\sigma_{\rm sys,trop}$), we used the simulations of \citet{Pradel2006} for both the VLBA and EVN measurements, as appropriate for our target-phase calibrator angular separation and target declination. To this, we added in quadrature a conservative estimate of the ionospheric systematics \citep{Reid2017}, taken as $\sigma_{\rm sys,ion} = 50\,\mu{\rm as}\left(\frac{\nu}{6.7\,\mathrm{GHz}}\right)^{-2}\left(\frac{\theta}{1^{\circ}}\right)$,
where $\nu$ is the observing frequency and $\theta$ is the angular separation between target and phase reference calibrator. The rms of the position of our check source (J1821 for BM467A, BM467O, BM467R and J1813 for the remaining data sets) was in all cases less than the conservative upper limits calculated from the combined effect of $\sigma_{\rm sys,trop}$ and $\sigma_{\rm sys,ion}$.

\section{Results and Analysis}\label{Results_analysis}
\subsection{A Bayesian approach for parallax fitting}\label{Results}
\begin{figure*}
\centering
\includegraphics[width=0.43\textwidth]{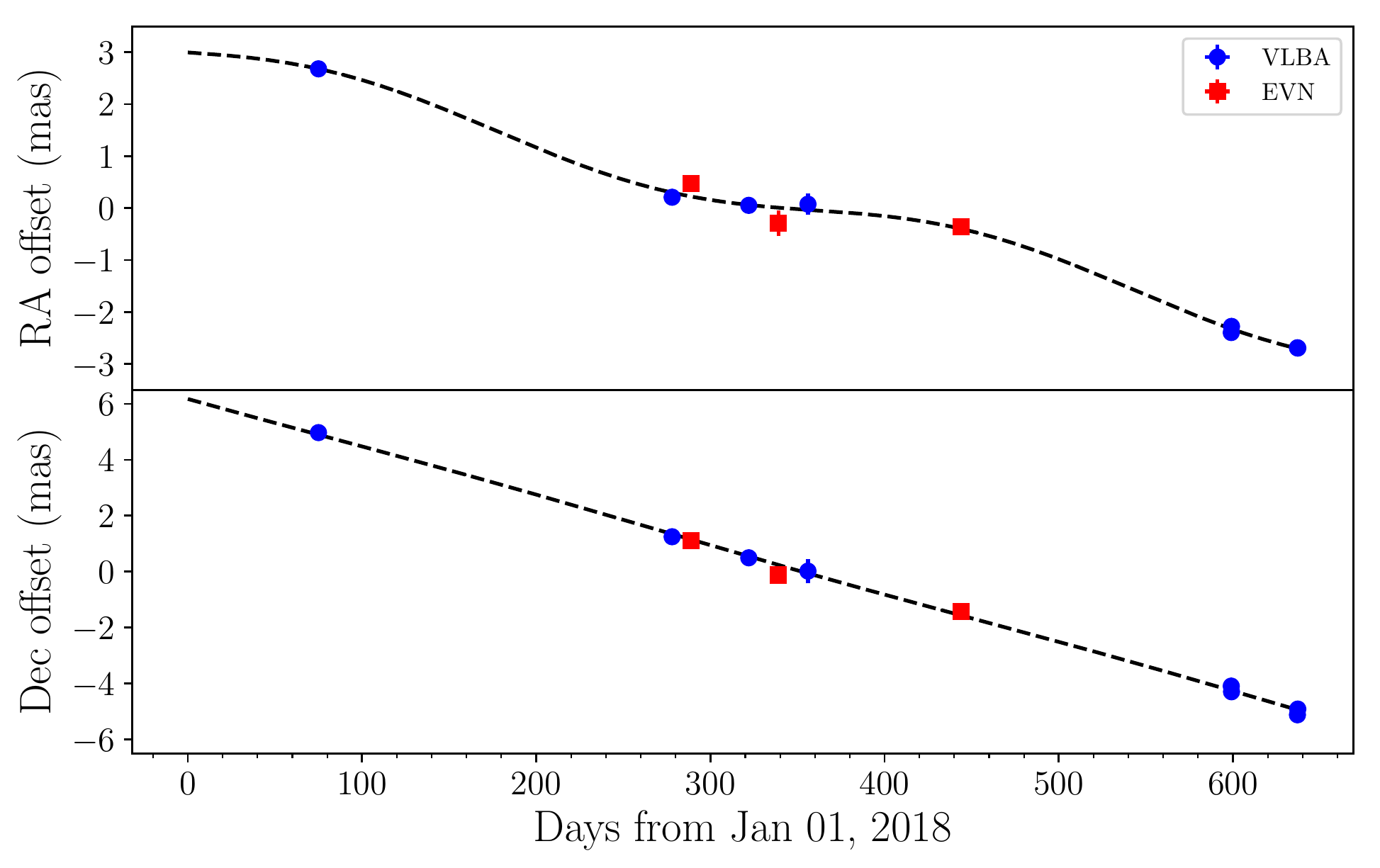}
\includegraphics[width=0.44\textwidth]{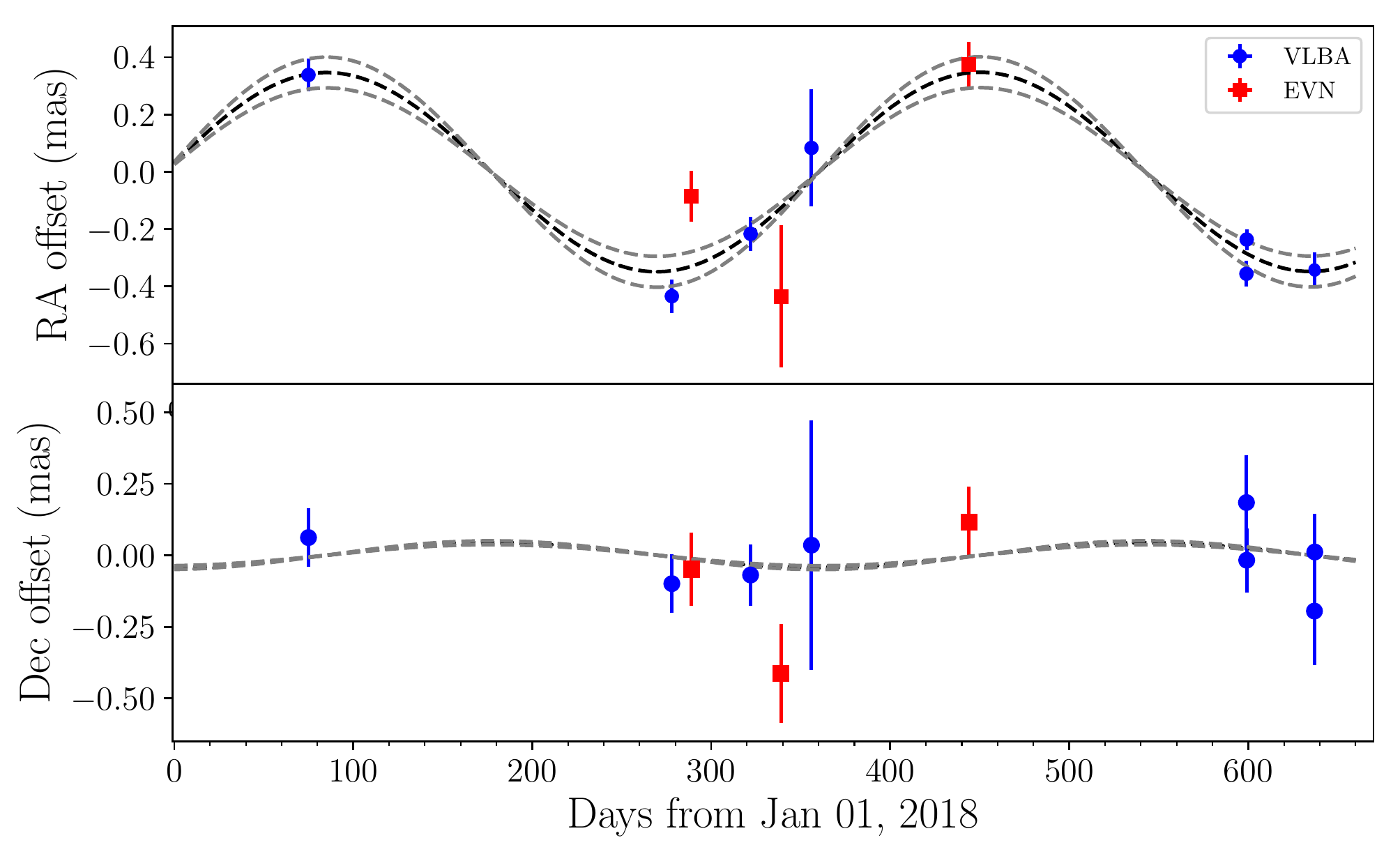}
\caption{Astrometry of MAXI\,J1820, as phase referenced to J1821 at 5\,GHz. The blue circles and red squares are the positions measured by the VLBA and the EVN, respectively. All marked positions have been corrected for the frame shift due to different calibrators and frequencies. The errors bars denote the statistical and systematic errors added in quadrature. Left panel: Motion in the plane of the sky relative to the fitted reference position, overlaid with the trajectory given by the best fitting proper motion and parallax (black dashed line). Right panel: Parallax signature of 0.348 (black dashed line) $\pm$ 0.033 (grey dashed lines)\,mas isolated by removing our best fitting proper motion.}
\label{propermotion}
\end{figure*}

To derive the proper motion ($\mu_{\alpha}\mathrm{cos}\delta$, $\mu_{\delta}$), parallax ($\pi$) and reference position (RA$_{0}$, Dec$_{0}$) of MAXIJ1820, we fit the position as a function of time \citep[e.g.][]{Loinard2007}, with a reference date equivalent to the midpoint of the observing campaign (MJD 58474). To perform the fit, we adopted a Bayesian approach, using the PYMC3 python package \citep{Salvatier2016} to implement a Hamilton Markov Chain Monte Carlo \citep[MCMC,][]{Neal2012} technique with a No-U-Turn Sampler \citep[NUTS;][]{Hoffman2011}. The \textit{Gaia}--DR2 proper motion and parallax measurements were used as priors in the procedure. We had observations at 5 and 15\,GHz and so fit for any potential core shift in the calibrator peak emission from 5\,GHz to 15\,GHz ($\alpha_{s}$,$\delta_{s}$), and used a flat prior from -1\,mas to 1\,mas for both $\alpha_{s}$and $\delta_{s}$ . We checked for convergence in our MCMC using the Gelman-Rubin diagnostic \citep{Gelman1992}. The posterior distributions for all parameters are Gaussian, and the median and the 68\,per cent confidence interval of each of the fitted parameters are given in Table \ref{tab:table3}.  All error bars reported hereafter are at 68\,per cent confidence, unless otherwise stated. The resulting fits to the sky motion are shown in Figure \ref{propermotion}.\par
\begin{table}
        \begin{center}
                \caption{The results of the Bayesian fitting algorithm. Here we report the median and 68$\%$ confidence interval from the posterior distributions of the fitted parameters. \label{tab:table3}}
                \begin{tabular}{l r}
                \hline \hline
                        Parameter & Value\\ \hline 
                        RA$_{0}$  & 18$^{\rmn{h}}$20$^{\rmn{m}}$21$\fs$9384568 $~\pm$ 0.0000024\\
                        Dec$_{0}$ & 07$\degr$11$\arcmin$07$\farcs$1649624 $~\pm$ 0.0000680\\
                        $\mu_{\alpha}\cos\delta$ (mas\,yr$^{-1}$) & $-$3.051 $~\pm$ 0.046\\
                        $\mu_{\delta}$ (mas\,yr$^{-1}$)& $-$6.394 $~\pm$ 0.075\\
                        $\pi$ (mas) & 0.348 $~\pm$ 0.033\\
                        $\alpha_{s}$ (mas) & $-$0.04 $~\pm$ 0.04\\
                        $\delta_{s}$ (mas) & 0.04 $~\pm$ 0.08\\
                        \hline
                \end{tabular}
        \end{center}
\vspace{-0.3cm}
\end{table}
\subsection{Distance from parallax}
\begin{figure}
\centering
\includegraphics[width=0.4\textwidth]{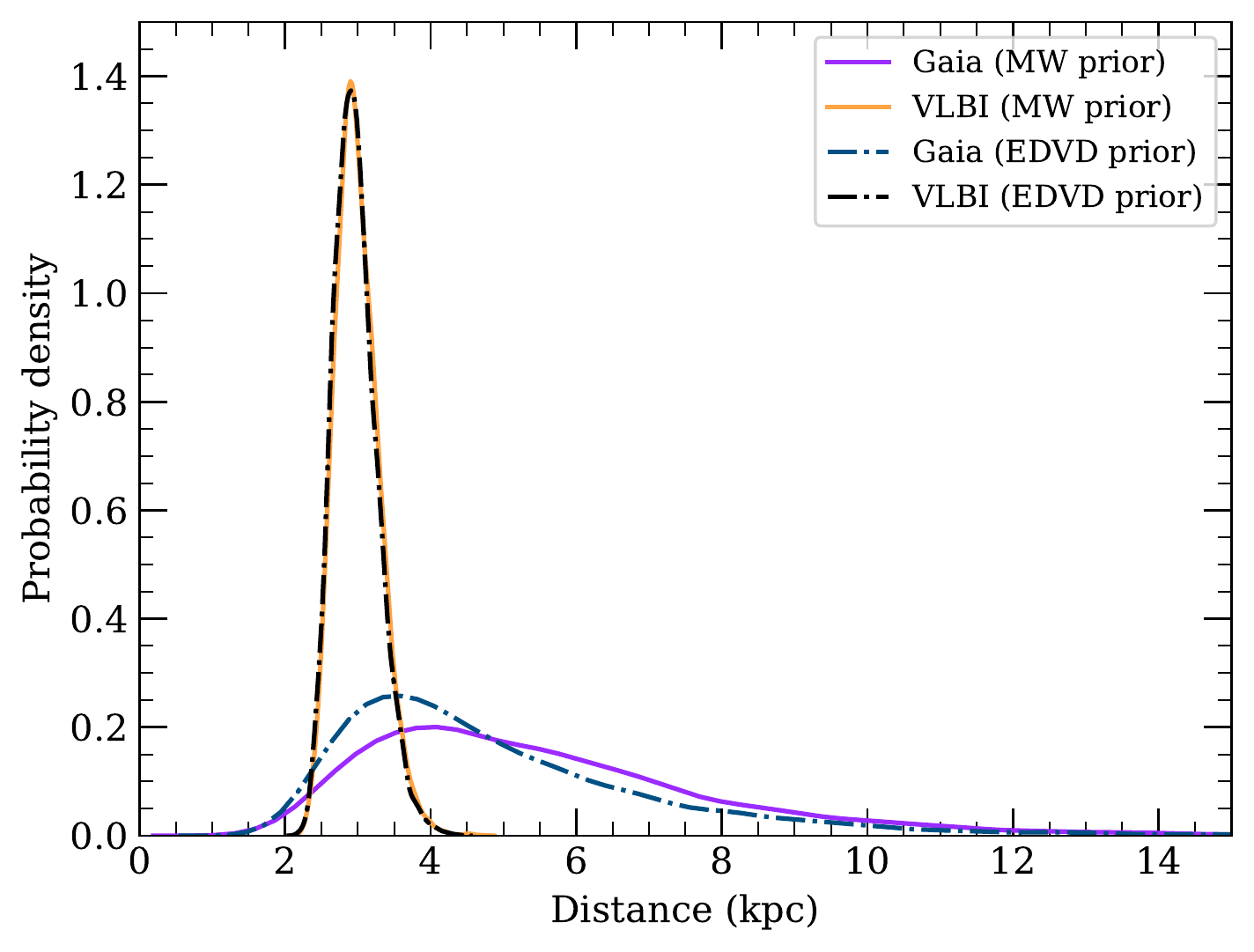}
\caption{Distance posterior distributions of MAXI\,J1820, for both our new VLBI and the \textit{Gaia}--DR2 parallax measurement. The \textit{Gaia}-DR2 distance posterior has a higher median value for the Milky Way (MW) prior (5.1$\pm$2.7\,kpc) than for the exponentially-decreasing volume density (EDVD) prior (4.4$\pm$2.4\,kpc), whereas our VLBI distance distribution (2.96$\pm$0.33\,kpc) is insensitive to the prior chosen.}
\label{distance}
\end{figure}
\vspace{-0.3cm}
Distances inferred by inverting low significance parallax measurements suffer from Lutz-Kelker bias and can thus be underestimated \citep{Lutz1973}. Thus the use of appropriate priors is essential to obtain meaningful distances \citep{Bailer-Jones2015,Astraatmadj2016}. We use a prior that models the X-ray binary density distribution in the Milky Way (MW prior hereon; \citet{Atri2019}) to estimate the distance posterior distribution for MAXI\,J1820. With our new and more precise parallax measurement (0.348$\pm$0.033\,mas) and the MW prior, we determined the distance posterior distribution (Figure~\ref{distance}), finding a median of 2.96\,kpc and a 68\,per cent confidence interval of $\pm$0.33\,kpc. We compared this distance posterior distribution to that obtained using the parallax measurement made by \textit{Gaia}-DR2, which gave a larger median value and a wider distribution ($5.1\pm2.7$)\,kpc. The mode and the 68\,per cent higher density interval values \citep{Bailer-Jones2015} of this distribution are 4.13$^{+2.57}_{-1.5}$\,kpc. We also tested an exponentially decreasing volume density (EDVD) prior \citep{Bailer-Jones2015} using the scaling length parameter L for BHXBs from \citet{Gandhi2019} and find that unlike the \textit{Gaia}--DR2 distance posterior distribution (median of $4.4\pm2.4$\,kpc; mode of 3.46$^{+2.13}_{-1.06}$\,kpc), our new distance measurement is unaffected by the chosen prior (see Figure \ref{distance}). This shows that the high significance of our parallax measurement has led to a model-independent distance to the source. 

\section{Discussion}\label{Discussion}
With our much improved distance measurement for MAXI\,J1820, we now consider the physical implications of our result.

\subsection{Transition and peak outburst luminosities}
The main outburst of MAXI\,J1820 was observed to have a peak X-ray flux of ($14\pm1)\times10^{-8}$\,erg\,cm$^{-2}$\,s$^{-1}$ in the 1--100\,keV band \citep{Shidatsu2019}. A distance of $2.96\pm0.33$\,kpc implies that the system only reached $0.15\pm0.03\,L_{\rm{Edd}}$ at the peak of its outburst, where $L_{\rm{Edd}}$ is the Eddington luminosity for a $10M_{\odot}$ BH (see Section \ref{bhmass}). The soft-to-hard state transition luminosity, from a flux of ($2.5\pm0.4)\times10^{-8}$\,erg\,cm\,$^{-2}$\,s$^{-1}$, implies that the system made this transition at $3\pm1$\,per cent$\,L_{\rm{Edd}}$, in agreement with the average luminosity fraction of 1.58$\pm$0.93\,per cent\,$L_{\rm{Edd}}$ for BHXBs \citep{Vahdat2019,Maccarone2003}. Thus, for MAXI\,J1820 the state transition luminosity provides a good distance indicator.

\subsection{Jet parameters}
Assuming intrinsic symmetry, the measured proper motions of corresponding approaching and receding jet ejecta along with a distance can be used to uniquely determine the jet speed and inclination angle between the jet and the line of sight \citep{Mirabel1994,Fender1999}, via $\beta\, \rm{cos}\,\textit{i} = \rm{\frac{\mu_{app}-\mu_{rec}}{\mu_{app}+\mu_{rec}}};\, \rm{tan\,\textit{i} = \frac{2\textit{d}}{c}\frac{\mu_{app}\, \mu_{rec}}{\mu_{app}-\mu_{rec}}}$
where $\beta$ is the velocity of the jet (normalised to the speed of light $c$), $\mu_{\mathrm{app}}$ and $\mu_{\mathrm{rec}}$ are the proper motions of the approaching and receding components of the jet, $i$ is the inclination angle of the jet to the line of sight, and $d$ is the distance to the system. \citet{Bright2020} measured the proper motions of corresponding jet ejecta event during the transition phase of MAXI\,J1820 $\mu_{\rm{app}} = 77 \pm 1$\,mas\,d$^{-1}$, $\mu_{\rm{rec}} = 33\pm 1$\,mas\,d$^{-1}$. 
We use this with our distance constraint to estimate $\beta$ = $0.89\pm0.09$ and $\theta = (63\pm3)^{\circ}$. 
\subsection{Implications for BH mass}\label{bhmass}
\citet{Torres2019} conducted optical spectroscopy on MAXI\,J1820 and reported a mass function of $f(M)=\frac{\left(M_{1} \sin i\right)^{3}}{\left(M_{1}+M_{2}\right)^{2}}=5.18 \pm 0.15 M_{\odot}$,
where $f(M)$ is the mass function, $i$ is the inclination angle, and $M_{1}$ and $ M_{2}$ are the masses of the BH and its companion. Assuming a mass ratio $q \equiv M_{2}/M_{1} = 0.12$, they constrained the inclination angle to be $69^{\circ} < i < 77^{\circ}$, and the BH mass to be 7--8\,$M_{\odot}$. Using our derived inclination angle of $(63\pm3)^{\circ}$, we re-calculated the BH mass to be $(9.2\pm1.3)\,M_{\odot}$. \citet{Torres2019} suggested that the value $q=0.12$ needs confirmation, so we also calculated the mass of the BH for the full suggested range of $q$ (0.03--0.4) to be $(10\pm2)M_{\odot}$. We used this updated inclination angle ($i$) and distance ($D$) to estimate the inner disc radius in the high/soft state $R_{in} = 77.9\pm1.0 \frac{D}{\rm{3\,kpc}} \frac{\rm{cos}\,\textit{i}}{\rm{cos}\,60^{\circ}}\,\rm{km}$ \citet{Shidatsu2019}. Equating this to the inner most stable orbit of the BH, we suggest that the BH in MAXI\,J1820 is likely slowly spinning \citep{McClintock2013}.

\subsection{Potential kick velocity (PKV)}
The velocity distribution of BHXBs at Galactic plane crossing has been used to determine the potential kick the system might have received at the birth of the BH \citep{Atri2019}, which is an indicator of the birth mechanism of the BH \citep{Nelemans1999,Fragos2009,Janka2017}. A robust PKV distribution requires good constraints on the proper motion, systemic radial velocity and the distance to the system. We used the parallax and proper motion measured in this work, combined with the systemic radial velocity of ($-21.6\pm 2.3$)\,km\,s$^{-1}$ \citep{Torres2019} to determine a PKV distribution with a median of 120\,km\,s$^{-1}$, and 5$^{\rm{th}}$ and 95$^{\rm{th}}$ percentiles of 95 and 150\,km\,s$^{-1}$, respectively. This velocity is higher than the typical velocity dispersion of stars in the Galaxy \citep[50\,km\,s$^{-1}$;][]{Mignard2000} and suggests that the system likely received a strong kick at birth, consistent with formation in a supernova explosion. 

\section{CONCLUSIONS}
We report a precise VLBI parallax measurement of (0.348$\pm$0.033)\,mas to the BHXB MAXI\,J1820. Using this parallax and a Bayesian prior, we inferred a distance of (2.96$\pm$0.33)\,kpc. We showed that MAXI\,J1820 reached ($15\pm3$) per cent L$_{\rm{Edd}}$ at the peak of its outburst. We constrained the jet inclination angle and velocity to be ($63\pm3)^{\circ}$ and ($0.89\pm0.09)c$, respectively. We also report an updated BH mass estimate of ($9.2\pm1.3)\,M_{\odot}$, and suggest the BH is slowly spinning and likely received a strong natal kick.

\section{Acknowledgements}
We would like to thank J.S. Bright for allowing us to use their work prior to publication that helped us uniquely solve for the jet parameters in Section 4.2. The National Radio Astronomy Observatory is a facility of the National Science Foundation operated under cooperative agreement by Associated Universities, Inc.  The European VLBI Network is a joint facility of independent European, African, Asian, and North American radio astronomy institutes. Scientific results from data presented in this publication are derived from the following EVN project code(s): EA062. e-VLBI research infrastructure in Europe is supported by the European Union's Seventh Framework Programme (FP7/2007-2013) under grant agreement number RI-261525 NEXPReS. JCAM-J is the recipient of an Australian Research Council Future Fellowship (FT140101082), funded by the Australian government. PGJ acknowledges funding from the European Research Council under ERC Consolidator Grant agreement no 647208. GRS acknowledges support from an NSERC Discovery Grant (RGPIN-06569-2016). DA acknowledges support from the Royal Society. TDR acknowledges support from a Netherlands Organisation for Scientific Research (NWO) Veni Fellowship. VT is supported by programme Laplas VI of the Romanian National Authority for Scientific Research.

\vspace{-0.3cm}



\bibliographystyle{mnras}
\bibliography{mj1820} 



\bsp	
\label{lastpage}
\end{document}